\documentstyle[amssymb,12pt,thmsa,sw20lart]{article}
\input{tcilatex}
\begin{document}

\title{Reduction of multipartite qubit density matrixes to bipartite qubit density
matrixes and criteria of partial separability of multipartite qubit density
matrixes}
\author{Zai-Zhe Zhong$^1$ \\
%EndAName
1. Department of Physics, Liaoning Normal University, Dalian 116029, \\
Liaoning, China.}
\maketitle

\begin{abstract}
The partial separability of multipartite qubit density matrixes is strictly
defined. We give a reduction way from N-partite qubit density matrixes to
bipartite qubit density matrixes, and prove a necessary condition that a
N-partite qubit density matrix to be partially separable is its reduced
density matrix to satisfy PPT condition.

PACC numbers: 03.67.Mn; 03.65.Ud; 03.67.Hk
\end{abstract}

Recently, an important task in modern quantum mechanics and quantum
information is to find the criteria of separability of density matrixes. The
first important result is the well-known positive partial transposition
(PPT, Peres-Horodecki) criteria[1,2$]$ for $2\times 2$ and $2\times 3$
systems. There are many studies about the criteria of separability for the
multipartite systems, see [3-8$].$

Generally, the common so-called `separability', in fact, is the
full-separability. For multipartite systems the problems are more complex,
there yet is other concept of separability weaker than full-separability,
i.e. the `partial separability'$,$ e.g. the A-BC-separability,
B-AC-separability for a tripartite qubit pure-state $\rho _{ABC}[8]$, etc..
Related to Bell-type inequalities and some criteria of partial separability
of multipartite systems, etc., see [9-12$].$ However, we yet need to
stricter define the concept of partial separability and find the simpler
criteria. In this paper, first we discuss how to define strictly the concept
of the partial separability corresponding to a partition. Next, we give a
new way that an arbitrary N-partite (N$\geq 3)$ qubit density matrix always
can be reduced in one step through to a bipartite qubit density matrix .
Thus, we prove an effective criterion: A necessary condition of a N-partite
qubit density matrix to be partially separable\ with respect to a partition
is that the corresponding reduced bipartite qubit density matrix is
separable, i.e. it satisfies\ the PPT condition. Some examples are given.

Suppose that $\rho _{i_1i_2\cdots i_N}$ is a density matrix for N-partite
qubit Hilbert space $H=\otimes _{s=1}^NH_s$ $,$ of which the standard basis
is $\left\{ \otimes _{s=1}^N\mid i_s>\right\} (i_s=0,1).$ Let ${\Bbb Z}_N$
be the integer set $\left\{ 1,2,\cdots ,N\right\} .$ If two subsets $\left(
r\right) _P\equiv \left\{ r_1,\cdots ,r_P\right\} $ and $\left( s\right)
_{N-P}\equiv \left\{ s_1,\cdots ,s_{N-P}\right\} $ in ${\Bbb Z}_N$ obey 
\begin{eqnarray}
\;1 &\leqslant &r_1<\cdots <r_P<N,\;1<s_1<\cdots <s_{N-P}\leqslant N 
\nonumber \\
\;\left( r\right) _P\cup \left( s\right) _{N-P} &=&{\Bbb Z}_N,\;\left(
r\right) _P\cap \left( s\right) _{N-P}=\emptyset (1\leqslant P<N)\;
\end{eqnarray}
where $P$ is an integer, 1$\leqslant P\leqslant N-1,$ the set $\left\{
\left( r\right) _P,\left( s\right) _{N-P}\right\} $ forms a partition of $%
{\Bbb Z}_N$, in the following we simply call it a `partition', and for the
sake of stress we denote it by symbol $\left( r\right) _P\Vert \left(
s\right) _{N-P}$. A partition $\left( r\right) _P\Vert \left( s\right)
_{N-P} $ corresponds to a permutation $S_{\left( r\right) _P\Vert \left(
s\right) _{N-P}}\equiv \left( 
\begin{array}{cccccc}
1, & \cdots & P, & P+1, & \cdots & N \\ 
r_1, & \cdots & r_P, & s_1, & \cdots & s_{N-P}
\end{array}
\right) $, by which a new matrix $\rho _{\left( r\right) _P\Vert \left(
s\right) _{N-P}}$ from $\rho _{i_1i_2\cdots i_N}$ is defined now, whose
entries are 
\begin{equation}
\left[ \rho _{\left( r\right) _P\Vert \left( s\right) _{N-P}}\right]
_{j_1\cdots j_N,\;k_1\cdots k_N}=\left[ \rho \right] _{j_{r_1}\cdots
j_{r_P}j_{s_1}\cdots j_{s_{N-P}},\;k_{r_1}\cdots k_{r_P}k_{s_1}\cdots
k_{s_{N-P}}}
\end{equation}
For instance, $\rho _{A\Vert BCD}=\rho _{AB\Vert CD}\;=\rho _{ABC\Vert
D}=\rho _{ABCD}$, and $\left[ \rho _{AC\Vert BD}\right] _{ijkl,rstu}=\left[
\rho _{ABCD}\right] _{ikjl,\;rtsu},$ $\left[ \rho _{C\Vert ABD}\right]
_{ijkl,rstu}=\left[ \rho _{ABCD}\right] _{kijl,.trsu}$, etc.. Generally, $%
\rho _{\left( r\right) _P\Vert \left( s\right) _{N-P}}\neq \rho
_{i_1i_2\cdots i_N}$, unless $\left( r\right) _P\Vert \left( s\right) _{N-P}$
just maintains the natural order of ${\Bbb Z}_N$ (i.e. $\left( r\right)
_P=\left( 1,\cdots ,P\right) ,\left( s\right) _{N-P}=\left( P+1,\cdots
,N\right) )$, then $\rho _{\left( r\right) _P\Vert \left( s\right)
_{N-P}}=\rho _{i_1i_2\cdots i_N}.$

{\bf Lemma. }For any{\bf \ }partition $\left( r\right) _P\Vert \left(
s\right) _{N-P},$ $\rho _{\left( r\right) _P\Vert \left( s\right) _{N-P}}$
is still a N-partite qubit density matrix.

{\bf Proof. }We only consider the case of tripartite qubit, the general
cases are completely similar (also see [11$]).$ Notice the permutation $%
S_{B\Vert AC}$ , then we have 
\begin{equation}
\;\rho _{B\Vert AC}=S\rho _{ABC}S^{\dagger },\;S=\left[ 
\begin{array}{cccccccc}
1 &  &  &  &  &  &  &  \\ 
& 1 &  &  &  &  &  &  \\ 
&  & 0 & 0 & 1 & 0 &  &  \\ 
&  & 0 & 0 & 0 & 1 &  &  \\ 
&  & 1 & 0 & 0 & 0 &  &  \\ 
&  & 0 & 1 & 0 & 0 &  &  \\ 
&  &  &  &  &  & 1 &  \\ 
&  &  &  &  &  &  & 1
\end{array}
\right]
\end{equation}
$S$ is an unitary matrix, therefore $\rho _{B\Vert AC}$ is still a
tripartite qubit density matrix. $\square $

Now, we consider how to more strictly define the partial separability.
Obviously, if a partition $\left( r\right) _P\Vert \left( s\right) _{N-P}$
maintains the natural order of ${\Bbb Z}_N$ (i.e. $\left( r\right) _P=\left(
1,2,\cdots ,P\right) ,$ $\left( s\right) _{N-P}=\left( P+1,P+2,\cdots
,N\right) ),$ then $\rho _{\left( r\right) _P\Vert \left( s\right)
_{N-P}}=\rho _{i_1i_2\cdots i_N}$ under the standard basis $\left\{ \otimes
_{s=1}^N\mid i_s>\right\} $, now the $\left( r\right) _P-\left( s\right)
_{N-P}$-separability can naturally be defined as that if $\rho
_{i_1i_2\cdots i_N}$ can be decomposed as $\rho _{\left( r\right) _P\Vert
\left( s\right) _{N-P}}=\rho _{i_1i_2\cdots i_N}=\sum\limits_\alpha p_\alpha
\rho _{\alpha ,\left( r\right) _P}\otimes \rho _{\alpha ,\left( s\right)
_{N-P}}$ with probabilities $p_\alpha $, where $\rho _{\alpha ,\left(
r\right) _P}$ and $\rho _{\alpha ,\left( s\right) _{N-P}}$, respectively,
are a $P$-partite and a $\left( N-P\right) $-partite qubit density matrixes
acting upon $\otimes _{m=1}^PH_m$ and $\otimes _{n=1}^{N-P}H_n$ for all $%
\alpha ,$ then we call $\rho _{i_1i_2\cdots i_N}$ to be $\left( r\right)
_P-\left( s\right) _{N-P}$-separable$.$ However, if the natural order of $%
{\Bbb Z}_N$ has been broken in $\left( r\right) _P\Vert \left( s\right)
_{N-P}$ $($ i.e. $s_1<r_P),$ then generally $\rho _{\left( r\right) _P\Vert
\left( s\right) _{N-P}}\neq \rho _{i_1i_2\cdots i_N}$ , the case is
different from the above. For instance, we consider a normalized pure-state $%
\rho _{ABCD}=\mid \Psi _{ABCD}><\Psi _{ABCD}\mid ,$ $\mid \Psi _{ABCD}>\in
H_A\otimes H_B\otimes H_C\otimes H_D$ of four spin-$\frac 12$ particles A,
B, C and D. Now, assume that $\mid \Psi _{ABCD}>$ has a special form as $%
\mid \Psi _{ABCD}>=\sum\limits_{i,j,k,l=0,1}c_{ik}c_{jl}\mid i_A>\otimes
\mid j_B>\otimes \mid k_C>\otimes \mid l_D>,$ where $c_{ik},c_{jl}\in {\Bbb C%
}^1$. If we keep up to use the original standard basis$,$ then we cannot
directly see the partial separability, because this choice of basis is
unsuitable. If we choose other nature basis $\left\{ \mid i_A>\otimes \mid
k_C>\otimes \mid j_B>\otimes \mid l_D>\right\} $ $($ this, in fact, means
that we are using $\rho _{AC\Vert BD}),$ under which we can consider the
state $\mid \Psi _{ACBD}^{\prime }>=\mid \Psi _{AC}>\otimes \mid \Psi _{BD}>$
, where $\mid \Psi _{AC}>=\sum\limits_{i,k=0,1}c_{ik}\mid i_A>\otimes \mid
k_C>,$ $\mid \Psi _{BD}>=\sum\limits_{j,l=0,1}c_{jl}\mid j_B>\otimes \mid
l_D>.$ Now, $\rho _{AC\Vert BD}=\rho _{AC}\otimes \rho _{BD}$, where $,$ $%
\rho _{AC}=\mid \Psi _{AC}><\Psi _{AC}\mid ,\rho _{BD}=\mid \Psi _{BD}><\Psi
_{BD}\mid $. $\mid \Psi _{ABCD}>$ and $\mid \Psi _{ABCD}>$, in fact, are the
same in physics, therefore to call $\rho _{ABCD}$ AC-BD-separable is
completely reasonable. Similarly, for the rest. Generalize to the cases of
mixed-states, thus we can generally define the concept of partial
separability as follows.

{\bf Definition. }{\it For the partition} $\left( r\right) _P\Vert \left(
s\right) _{N-P}$ , {\it a N-partite qubit density matrix }$\rho
_{i_1i_2\cdots i_N}$ {\it acting} {\it upon H}$=\otimes _{s=1}^NH_s$ {\it is}
{\it called to be }$\left( r\right) _P-\left( s\right) _{N-P}${\it %
-separable if the corresponding density matrix }$\rho _{\left( r\right)
_P\Vert \left( s\right) _{N-P}}${\it \ can be decomposed as } 
\begin{equation}
\rho _{\left( r\right) _P\Vert \left( s\right) _{N-P}}=\sum\limits_\alpha
p_\alpha \rho _{\alpha ,\left( r\right) _P}\otimes \rho _{\alpha ,\left(
s\right) _{N-P}}
\end{equation}
{\it where }$\rho _{\alpha ,\left( r\right) _P}${\it \ and }$\rho _{\alpha
,\left( s\right) _{N-P}}${\it , respectively,} {\it are a }$P${\it -partite
and a }$\left( N-P\right) ${\it -partite qubit density matrixes acting upon }%
$\otimes _{m=1}^PH_{r_m}${\it \ and }$\otimes _{n=1}^{N-P}H_{s_n}$ {\it for
all }$\alpha ,$ {\it and }$0<p_\alpha \leq 1,\;\sum\limits_\alpha p_\alpha
=1.$ {\it If }$\rho _{i_1i_2\cdots i_N}${\it \ is not }$\left( r\right)
_P-\left( s\right) _{N-P}${\it -separable, then we call it }$\left( r\right)
_P-\left( s\right) _{N-P}${\it -inseparable.}

For the distinct partitions $\rho _{i_1i_2\cdots i_N}$ can have distinct
separability. Of course, if a $\rho _{i_1i_2\cdots i_N}$ is partially
inseparable for some partition, then it must be entangled. Here, in passing,
we point out that how to find the general relations between the partial
separability and the ordinary separability (full-separability), generally,
is not a simple problem. For instance, we can make such a multipartite qubit
density matrix $\stackrel{\backsim }{\rho }$ $($similar to the theorem 1 in
[13,14]), and by using of the technique in this paper, we can prove that $%
\stackrel{\backsim }{\rho }$ always is partially separable for all possible
partitions $\left( r\right) _P\Vert \left( s\right) _{N-P}\left( 1\leqslant
P\leqslant N-1\right) ,$ but $\stackrel{\backsim }{\rho }$ is entangled (not
full-separability).

In order to find the criteria of \ partial separability, first we discuss
how to reduce a multipartite qubit density matrix in one step through to a
bipartite qubit density matrix. For a given partition $\left( r\right)
_P\Vert \left( s\right) _{N-P},$ let two sets $\left( r\right) _P$ and $%
\left( s\right) _{N-P},$ respectively be separated again as follows, 
\begin{eqnarray}
\left( r^{\prime }\right) _{P^{\prime }} &=&\left\{ r_1^{\prime },\cdots
,r_{P^{\prime }}^{\prime }\right\} ,\left( r^{\prime \prime }\right)
_{P^{\prime \prime }}=\left\{ r_1^{\prime \prime },\cdots ,r_{P^{^{\prime
\prime }}}^{\prime \prime }\right\} \text{, one of them can be the null set}
\nonumber \\
\left( s^{\prime }\right) _{Q^{\prime }} &=&\left\{ s_1^{\prime },\cdots
,s_{Q^{\prime }}^{\prime }\right\} ,\left( s^{\prime \prime }\right)
_{Q^{\prime \prime }}=\left\{ s_1^{\prime \prime },\cdots ,s_{Q^{\prime
\prime }}^{\prime \prime }\right\} \text{, one of them can be the null set} 
\nonumber \\
\text{ }r_1^{\prime } &<&r_2^{\prime }<\cdots <r_{P^{\prime }}^{\prime
},\;\;r_1^{\prime \prime }<r_2^{\prime \prime }<\cdots \;<r_{P^{\prime
\prime }}^{\prime \prime } \\
\;s_1^{\prime } &<&\text{ }s_2^{\prime }<\cdots <s_{Q^{\prime }}^{\prime
},\;\;s_1^{\prime \prime }<\text{ }s_2^{\prime \prime }<\cdots <s_{Q^{\prime
\prime }}^{\prime \prime }\text{ }  \nonumber \\
\left( r\right) _P &=&\left( r^{\prime }\right) _{P^{\prime }}\cup \left(
r^{\prime \prime }\right) _{P^{\prime \prime }},\;\left( r^{\prime }\right)
_{P^{\prime }}\cap \left( r^{\prime \prime }\right) _{P^{\prime \prime
}}=\emptyset (0\leq P^{\prime },P^{\prime \prime }\leq P\text{ and }%
P^{\prime }+P^{\prime \prime }=P)  \nonumber \\
\;\left( s\right) _{N-P} &=&\left( s^{\prime }\right) _{Q^{\prime }}\cup
\left( s^{\prime \prime }\right) _{Q^{\prime \prime }},\left( s^{\prime
}\right) _{Q^{\prime }}\cap \left( s^{\prime \prime }\right) _{Q^{\prime
\prime }}=\emptyset \left( 0\leq Q^{\prime },Q^{\prime \prime }\leq N-P\text{
and }Q^{\prime }+Q^{\prime }=N-P\right)  \nonumber
\end{eqnarray}
now we rewrite the partition added these partitions as $\left[ \left(
r^{\prime }\right) _{P^{\prime }},\left( r^{\prime \prime }\right)
_{P^{\prime \prime }}\right] \Vert \left[ \left( s^{\prime }\right)
_{Q^{\prime }},\left( s^{\prime \prime }\right) _{Q^{\prime \prime }}\right] 
$. Now we define the matrix $\rho _{\left[ \left( r^{\prime }\right)
_{P^{\prime }},\left( r^{\prime \prime }\right) _{P^{\prime \prime }}\right]
\Vert \left[ \left( s^{\prime }\right) _{m-P^{\prime }},\left( s^{\prime
\prime }\right) _{m-P^{\prime \prime }}\right] }$ by 
\begin{eqnarray}
&&\rho _{\left[ \left( r^{\prime }\right) _{P^{\prime }},\left( r^{\prime
\prime }\right) _{P^{\prime \prime }}\right] \Vert \left[ \left( s^{\prime
}\right) _{m-P^{\prime }},\left( s^{\prime \prime }\right) _{m-P^{\prime
\prime }}\right] }\text{ = the submatrix in }\rho _{i_1\cdots i_N}\text{
consisting of all entries }  \nonumber \\
&&\text{with form as }\left[ \rho \right] _{x_1x_2\cdots x_N,\;y_1y_2\cdots
y_N}
\end{eqnarray}
which must be a 4$\times 4$ matrix, where the values of $x_k$ and $y_k\left(
k=1,\cdots ,N\right) ,$ respectively, are determined by 
\begin{eqnarray}
x_k &=&i\text{ for }k\in \left( r^{\prime }\right) _{P^{\prime }},\;x_k=1-i%
\text{ for }k\in \left( r^{\prime \prime }\right) _{P^{\prime \prime }} 
\nonumber \\
\;x_k &=&j\text{ for }k\in \left( s^{\prime }\right) _{Q^{\prime }},\;x_k=1-j%
\text{ for }k\in \left( s^{\prime \prime }\right) _{Q^{\prime \prime }} 
\nonumber \\
y_k &=&u\text{ for }k\in \left( r^{\prime }\right) _{P^{\prime }},\;y_k=1-u%
\text{ for }k\in \left( r^{\prime \prime }\right) _{P^{\prime \prime }} \\
\;y_k &=&v\text{ for }k\in \left( s^{\prime }\right) _{Q^{\prime }},\;y_k=1-v%
\text{ for }k\in \left( s^{\prime \prime }\right) _{Q^{\prime \prime }}\; 
\nonumber
\end{eqnarray}
where $i,j,u,v=0,1$. E.g. 
\begin{eqnarray}
\rho _{\left[ \left( AC\right) ,\emptyset \right] \Vert \left[ \left(
B\right) ,\left( D\right) \right] } &=&\text{the submatrix in }\rho _{ABCD}%
\text{ consisting of all entries with}  \nonumber \\
\text{ form as }\left[ \rho \right] _{iji\left( 1-j\right) ,uvu\left(
1-v\right) } &=&\left[ 
\begin{array}{llll}
\left[ \rho \right] _{0001,0001} & \left[ \rho \right] _{0001,0100} & \left[
\rho \right] _{0001,1011} & \left[ \rho \right] _{0001,1110} \\ 
\left[ \rho \right] _{0100,0001} & \left[ \rho \right] _{0100,0100} & \left[
\rho \right] _{0100,1011} & \left[ \rho \right] _{0100,1110} \\ 
\left[ \rho \right] _{1011,0001} & \left[ \rho \right] _{1011,0100} & \left[
\rho \right] _{1011,1011} & \left[ \rho \right] _{1011,1110} \\ 
\left[ \rho \right] _{1110,0001} & \left[ \rho \right] _{1110,0100} & \left[
\rho \right] _{1110,1011} & \left[ \rho \right] _{1110,1110}
\end{array}
\right]
\end{eqnarray}
etc.. Now we define the 4$\times 4$ matrix {\it \ }$\rho _{\left( \left(
r\right) _P-\left( s\right) _{N-P}\right) }$ by 
\begin{equation}
\rho _{\left( \left( r\right) _P-\left( s\right) _{N-P}\right) }=\sum_{\Sb 
\text{ for all possible }\left[ \left( r^{\prime }\right) _{P^{\prime
}},\left( r^{\prime \prime }\right) _{P^{\prime \prime }}\right] \Vert
\left[ \left( s^{\prime }\right) _{Q^{\prime }},\left( s^{\prime \prime
}\right) _{Q^{\prime \prime }}\right] ,\;  \\ \text{and }\rho _{\left[
\left( r^{\prime }\right) _{P^{\prime }},\left( r^{\prime \prime }\right)
_{P^{\prime \prime }}\right] \Vert \left[ \left( s^{\prime }\right)
_{m-P^{\prime }},\left( s^{\prime \prime }\right) _{m-P^{\prime \prime
}}\right] }\text{ are not repeated}  \endSb }\rho _{\left[ \left( r^{\prime
}\right) _{P^{\prime }},\left( r^{\prime \prime }\right) _{P^{\prime \prime
}}\right] \Vert \left[ \left( s^{\prime }\right) _{m-P^{\prime }},\left(
s^{\prime \prime }\right) _{m-P^{\prime \prime }}\right] }\text{ }
\end{equation}
where we notice that there are indeed repeated{\em \ }$\rho _{\left[ \left(
r^{\prime }\right) _{P^{\prime }},\left( r^{\prime \prime }\right)
_{P^{\prime \prime }}\right] \Vert \left[ \left( s^{\prime }\right)
_{m-P^{\prime }},\left( s^{\prime \prime }\right) _{m-P^{\prime \prime
}}\right] }${\em , }in fact, {\em \ }$\rho _{\left[ \left( r^{\prime
}\right) _{P^{\prime }},\left( r^{\prime \prime }\right) _{P^{\prime \prime
}}\right] \Vert \left[ \left( s^{\prime }\right) _{m-P^{\prime }},\left(
s^{\prime \prime }\right) _{m-P^{\prime \prime }}\right] }$

$=\rho _{\left[ \left( r^{\prime }\right) _{P^{\prime }},\left( r^{\prime
\prime }\right) _{P^{\prime \prime }}\right] \Vert \left[ \left( s^{\prime
\prime }\right) _{m-P^{\prime \prime }},\left( s^{\prime }\right)
_{m-P^{\prime }}\right] }$

=$\rho _{\left[ \left( r^{\prime \prime }\right) _{P^{\prime \prime
}},\left( r^{\prime }\right) _{P^{\prime }}\right] \Vert \left[ \left(
s^{\prime }\right) _{m-P^{\prime }},\left( s^{\prime \prime }\right)
_{m-P^{\prime \prime }}\right] }${\em \ }$=\rho _{\left[ \left( r^{\prime
\prime }\right) _{P^{\prime \prime }},\left( r^{\prime }\right) _{P^{\prime
}}\right] \Vert \left[ \left( s^{\prime \prime }\right) _{m-P^{\prime \prime
}},\left( s^{\prime }\right) _{m-P^{\prime }}\right] },$ etc.. For instance,
we have 
\begin{eqnarray}
\rho _{\left( A-BC\right) } &=&\rho _{\left[ \left( A\right) ,\emptyset
\right] \Vert \left[ \left( BC\right) ,\emptyset \right] }+\rho _{\left[
\left( A\right) ,\emptyset \right] \Vert \left[ \left( B\right) ,\left(
C\right) \right] }  \nonumber \\
\rho _{\left( B-ACD\right) } &=&\rho _{\left[ \left( B\right) ,\emptyset
\right] \Vert \left[ \left( ACD\right) ,\emptyset \right] }+\rho _{\left[
\left( B\right) ,\emptyset \right] \Vert \left[ \left( AC\right) ,\left(
D\right) \right] }+\rho _{\left[ \left( B\right) ,\emptyset \right] \Vert
\left[ \left( AD\right) ,\left( C\right) \right] }+\rho _{\left[ \left(
B\right) ,\emptyset \right] \Vert \left[ \left( A\right) ,\left( CD\right)
\right] }  \nonumber \\
\rho _{\left( AC-BD\right) } &=&\rho _{\left[ \left( AC\right) ,\emptyset
\right] \Vert \left[ \left( BD\right) ,\emptyset \right] }+\rho _{\left[
\left( AC\right) ,\emptyset \right] \Vert \left[ \left( B\right) ,\left(
D\right) \right] }+\rho _{\left[ \left( A\right) ,\left( C\right) \right]
\Vert \left[ \left( BD\right) ,\emptyset \right] }+\rho _{\left[ \left(
A\right) ,\left( C\right) \right] \Vert \left[ \left( B\right) ,\left(
D\right) \right] }  \nonumber \\
\rho _{\left( AC-BDE\right) } &=&\rho _{\left[ \left( AC\right) ,\emptyset
\right] \Vert \left[ \left( BDE\right) ,\emptyset \right] }+\rho _{\left[
\left( AC\right) ,\emptyset \right] \Vert \left[ \left( BD\right) ,\left(
E\right) \right] }+\rho _{\left[ \left( AC\right) ,\emptyset \right] \Vert
\left[ \left( BE\right) ,\left( D\right) \right] }  \nonumber \\
&&+\rho _{\left[ \left( AC\right) ,\emptyset \right] \Vert \left[ \left(
B\right) ,\left( DE\right) \right] }+\rho _{\left[ \left( A\right) ,\left(
C\right) \right] \Vert \left[ \left( BDE\right) ,\emptyset \right] }+\rho
_{\left[ \left( A\right) ,\left( C\right) \right] \Vert \left[ \left(
BD\right) ,\left( E\right) \right] } \\
&&+\rho _{\left[ \left( A\right) ,\left( C\right) \right] \Vert \left[
\left( BE\right) ,\left( D\right) \right] }+\rho _{\left[ \left( A\right)
,\left( C\right) \right] \Vert \left[ \left( B\right) ,\left( DE\right)
\right] }  \nonumber
\end{eqnarray}
etc..

As an example, the above reduction procedures from $\rho _{ABCD}$\ to $\rho
_{\left( AC-BD\right) }$ can be described as $\rho _{ABCD}\longrightarrow $\ 
$\rho _{\left( AC-BD\right) }=\rho _{\left[ \left( AC\right) ,\emptyset
\right] \Vert \left[ \left( BD\right) ,\emptyset \right] }+\rho _{\left[
\left( AC\right) ,\emptyset \right] \Vert \left[ \left( B\right) ,\left(
D\right) \right] }+\rho _{\left[ \left( A\right) ,\left( C\right) \right]
\Vert \left[ \left( BD\right) ,\emptyset \right] }+\rho _{\left[ \left(
A\right) ,\left( C\right) \right] \Vert \left[ \left( B\right) ,\left(
D\right) \right] }$

$\equiv \sigma _\vartriangle +\sigma _{\times }+\sigma _{\diamond }+\sigma
_{\wedge },$\ where the submatrixes $\sigma _\vartriangle ,\sigma _{\times
},\sigma _{\diamond }$ and $\sigma _{\wedge },$\ respectively, consist of
the entries `$\vartriangle $', `$\times $',`$\diamond $' and `$\wedge $' in $%
\rho _{ABCD}$\ as in the following figure ($\sigma _{\times }$ is just the
matrix in Eq.(8)) 
\begin{equation}
\begin{array}{lllllllllllllllll}
& ^{_{_{0000}}} & ^{_{_{0001}}} & ^{_{_{0010}}} & ^{_{_{0011}}} & 
^{_{_{0100}}} & ^{_{_{0101}}} & ^{_{_{0110}}} & ^{_{_{0111}}} & ^{_{_{1000}}}
& ^{_{_{1001}}} & ^{_{_{1010}}} & ^{_{_{1011}}} & ^{_{_{1100}}} & 
^{_{_{1101}}} & ^{_{_{1110}}} & ^{_{_{^{_{1111}}}}} \\ 
^{_{_{0000}}} & \bigtriangleup &  &  &  &  & \bigtriangleup &  &  &  &  & 
\bigtriangleup &  &  &  &  & \bigtriangleup \\ 
^{_{_{0001}}} &  & \times &  &  & \times &  &  &  &  &  &  & \times &  &  & 
\times &  \\ 
_{^{_{0010}}} &  &  & \diamond &  &  &  &  & \diamond & \diamond &  &  &  & 
& \diamond &  &  \\ 
_{^{_{0011}}} &  &  &  & \wedge &  &  & \wedge &  &  & \wedge &  &  & \wedge
&  &  &  \\ 
^{_{_{0100}}} &  & \times &  &  & \times &  &  &  &  &  &  & \times &  &  & 
\times &  \\ 
^{_{_{0101}}} & \bigtriangleup &  &  &  &  & \bigtriangleup &  &  &  &  & 
\bigtriangleup &  &  &  &  & \bigtriangleup \\ 
^{_{_{0110}}} &  &  &  & \wedge &  &  & \wedge &  &  & \wedge &  &  & \wedge
&  &  &  \\ 
_{^{_{0111}}} &  &  & \diamond &  &  &  &  & \diamond & \diamond &  &  &  & 
& \diamond &  &  \\ 
^{_{_{1000}}} &  &  & \diamond &  &  &  &  & \diamond & \diamond &  &  &  & 
& \diamond &  &  \\ 
_{^{_{1001}}} &  &  &  & \wedge &  &  & \wedge &  &  & \wedge &  &  & \wedge
&  &  &  \\ 
^{_{_{1010}}} & \bigtriangleup &  &  &  &  & \bigtriangleup &  &  &  &  & 
\bigtriangleup &  &  &  &  & \bigtriangleup \\ 
_{^{_{1011}}} &  & \times &  &  & \times &  &  &  &  &  &  & \times &  &  & 
\times &  \\ 
_{^{_{1100}}} &  &  &  & \wedge &  &  & \wedge &  &  & \wedge &  &  & \wedge
&  &  &  \\ 
^{_{_{1101}}} &  &  & \diamond &  &  &  &  & \diamond & \diamond &  &  &  & 
& \diamond &  &  \\ 
^{_{_{1110}}} &  & \times &  &  &  &  &  &  &  &  &  & \times &  &  & \times
&  \\ 
^{_{_{1111}}} & \bigtriangleup &  &  &  &  & \bigtriangleup &  &  &  &  & 
\bigtriangleup &  &  &  &  & \bigtriangleup
\end{array}
\end{equation}

Similarly, we can consider higher dimensional cases. As for the ordinary
bipartite qubit density matrix $\rho _{AB},$\ we can take $\rho _{\left(
A-B\right) }\equiv \rho _{AB}$.

Sum up, generally we can define the 4$\times 4$\ matrix $\rho _{\left(
\left( r\right) _P-\left( s\right) _{N-P}\right) }$\ for a given $\left(
r\right) _P\Vert \left( s\right) _{N-P}.$\ In addition,\ it is easily
verified that for any partition $\left( r\right) _P\Vert \left( s\right)
_{N-P},$\ $\rho _{\left( \left( s\right) _{N-P}-\left( r\right) _P\right) }$%
\ is the transposition of $\rho _{\left( \left( u\right) _P-\left( s\right)
_{N-P}\right) },$\ therefore from viewpoint of partial separability, we
don't have to distinguish between the partitions $\left( r\right) _P\Vert
\left( s\right) _{N-P}$\ and $\left( s\right) _{N-P}\Vert \left( r\right)
_P. $

{\bf Theorem 1. }{\it For any partition }$\left( r\right) _P\Vert \left(
s\right) _{N-P},${\it \ }$\rho _{\left( \left( r\right) _P-\left( s\right)
_{N-P}\right) }${\it \ is a bipartite qubit density matrix, therefore }$\rho
_{\left( \left( r\right) _P-\left( s\right) _{N-P}\right) }${\it , in fact,} 
{\it is a reduction of the N-partite qubit density matrix }$\rho
_{i_1i_2\cdots i_N}${\it . }

{\bf Proof. }The fact must proved only is that $\rho _{\left( \left(
r\right) _P-\left( s\right) _{N-P}\right) }$\ is surely a bipartite qubit
density matrix. Here we only discuss in detail the cases of quadripartite
qubit states, since the generalization is completely straightforward. In the
first place, we prove that the theorem holds for a pure-state $\rho _{ABCD}$%
. Suppose that $\rho _{ABCD}=\mid \Psi _{ABCD}><\Psi _{ABCD}\mid $ is a
normalized pure-state$,$ where $\;\mid \Psi
_{ABCD}>=\sum_{i,j,k,l=0,1}c_{ijkl}\mid i_A>\otimes \mid j_B>\otimes \mid
k_C>\otimes \mid l_D>,$ $\sum_{i,j,k,l=0,1}\left| c_{ijkl}\right| ^2=1$. Let 
\begin{eqnarray}
&\mid &\Phi _{\bigtriangleup }>=\sum_{i,j=0,1}c_{ijij}\mid i_x>\otimes \mid
j_y>,\;\mid \Phi _{\times }>=\sum_{i,j=0,1}c_{iji\left( 1-j\right) }\mid
i_x>\otimes \mid j_y> \\
&\mid &\Phi _{\diamond }>=\sum_{i,j=0,1}c_{ij\left( 1-i\right) j}\mid
i_x>\otimes \mid j_y>,\;\mid \Phi _{\wedge }>=\sum_{i,j=0,1}c_{ij\left(
1-i\right) j\left( 1-j\right) }\mid i_x>\otimes \mid j_y>  \nonumber
\end{eqnarray}
where $x$ and $y$ are form particles. Make normalization, we obtain $\rho
_{\bigtriangleup }=\mid \varphi _{\bigtriangleup }><\varphi _{\bigtriangleup
}\mid ,\;\mid \varphi _{\bigtriangleup }>=\eta _{\bigtriangleup }^{-1}\mid
\Phi _{\bigtriangleup }$ \TEXTsymbol{>},\ $\rho _{\times }=\mid \varphi
_{\times }><\varphi _{\times }\mid ,\;\varphi _{\times }=\eta _{\times
}^{-1}\mid \Phi _{\times }>,$ $\rho _{\diamond }=\mid \varphi _{\diamond
}><\varphi _{\diamond }\mid ,\;\mid \varphi _{\diamond }>=\eta _{\diamond
}^{-1}\mid \Phi _{\diamond }>,$ $\rho _{\wedge }=\mid \varphi _{\wedge
}><\varphi _{\wedge }\mid ,\;\mid \varphi _{\wedge }>=\eta _{\wedge
}^{-1}\mid \Phi _{\wedge }$ \TEXTsymbol{>}, where the normalization factors
are 
\begin{eqnarray}
\eta _{\bigtriangleup } &=&\sqrt{\sum_{i,j=0,1}\left| c_{ijij}\right| ^2}%
,\;\eta _{\times }=\sqrt{\sum_{i,j=0,1}\left| c_{iji\left( 1-j\right)
}\right| ^2}  \nonumber \\
\eta _{\diamond } &=&\sqrt{\sum_{i,j=0,1}\left| c_{ij\left( 1-i\right)
j}\right| ^2},\;\eta _{\wedge }=\sqrt{\sum_{i,j=0,1}\left| c_{ij\left(
1-i\right) \left( 1-j\right) }\right| ^2}
\end{eqnarray}
It can be directly verified that from Eq.(10) we have 
\begin{equation}
\rho _{\left( AC-BD\right) }=\eta _{\bigtriangleup }^2\rho _{\bigtriangleup
}+\eta _{\times }^2\rho _{\times }+\eta _{\diamond }^2\rho _{\diamond }+\eta
_{\wedge }^2\rho _{\wedge }
\end{equation}
where $\rho _{\bigtriangleup },$\ $\rho _{\times },\rho _{\diamond },\rho
_{\wedge }$ all are bipartite qubit pure-states. It is easily seen that
since $\mid \Psi _{ABCD}>$ is normalized, $\eta _{\bigtriangleup }^2+\eta
_{\times }^2+\eta _{\diamond }^2+\eta _{\wedge
}^2=\sum\limits_{i,j,k,l=0,1}\left| c_{ijkl}\right| ^2=1$. This means that $%
\rho _{\left( AC-BD\right) }$ is a bipartite qubit density matrix(a mixed
state) for this pur-state $\rho _{ABCD}$.

Secondly, if $\rho _{ABCD}=\sum\limits_\alpha p_\alpha \rho _{\alpha \left(
ABCD\right) }$ is a mixed-state, where every $\rho _{\alpha \left(
ABCD\right) }$ is a quadripartite qubit pure-state with probabilities $%
p_\alpha ,$ then from Eq.(10) we have $\rho _{\left( AC-BD\right)
}=\sum_\alpha p_{\alpha ,\;}\left( \rho _\alpha \right) _{\left(
AC-BD\right) }.$ Since every $\left( \rho _\alpha \right) _{\left(
AC-BD\right) }$ is a bipartite qubit density matrix$,$ $\rho _{\left(
AC-BD\right) }$ is a density matrix (a mixed-state).

A similar way can be extended to higher dimensional case, the key is that
when $\rho _{i_1,\cdots ,i_N}$ is a pure-state$,$ then $\rho _{\left[ \left(
r^{\prime }\right) _{P^{\prime }},\left( r^{\prime \prime }\right)
_{P^{\prime \prime }}\right] \Vert \left[ \left( s^{\prime }\right)
_{m-P^{\prime }},\left( s^{\prime \prime }\right) _{m-P^{\prime \prime
}}\right] }$

$=\mid \Psi _{\left[ \left( r^{\prime }\right) _{P^{\prime }},\left(
r^{\prime \prime }\right) _{P^{\prime \prime }}\right] \Vert \left[ \left(
s^{\prime }\right) _{m-P^{\prime }},\left( s^{\prime \prime }\right)
_{m-P^{\prime \prime }}\right] }><\Psi _{\left[ \left( r^{\prime }\right)
_{P^{\prime }},\left( r^{\prime \prime }\right) _{P^{\prime \prime }}\right]
\Vert \left[ \left( s^{\prime }\right) _{m-P^{\prime }},\left( s^{\prime
\prime }\right) _{m-P^{\prime \prime }}\right] },$where the pure-state 
\begin{eqnarray}
&\mid &\Psi _{\left[ \left( r^{\prime }\right) _{P^{\prime }},\left(
r^{\prime \prime }\right) _{P^{\prime \prime }}\right] \Vert \left[ \left(
s^{\prime }\right) _{m-P^{\prime }},\left( s^{\prime \prime }\right)
_{m-P^{\prime \prime }}\right] }>=\sum_{i,j=0,1}c_{x_1x_2\cdots x_N}\mid
x_1>\otimes \cdots \otimes \mid x_N>  \nonumber \\
&&\text{ (}x_1,x_2,\cdots ,x_N\text{ are determined by Eq.(}7\text{))}
\end{eqnarray}
therefore we just have 
\begin{eqnarray}
&\mid &\Psi _{i_1,\cdots ,i_N}> \\
&=&\sum_{\Sb \text{ for all possible }\left[ \left( r^{\prime }\right)
_{P^{\prime }},\left( r^{\prime \prime }\right) _{P^{\prime \prime }}\right]
\Vert \left[ \left( s^{\prime }\right) _{Q^{\prime }},\left( s^{\prime
\prime }\right) _{Q^{\prime \prime }}\right] ,\;  \\ \text{and }\mid \Psi
_{\left[ \left( r^{\prime }\right) _{P^{\prime }},\left( r^{\prime \prime
}\right) _{P^{\prime \prime }}\right] \Vert \left[ \left( s^{\prime }\right)
_{m-P^{\prime }},\left( s^{\prime \prime }\right) _{m-P^{\prime \prime
}}\right] }\text{ \TEXTsymbol{>} are not repeated}  \endSb }\mid \Psi
_{\left[ \left( r^{\prime }\right) _{P^{\prime }},\left( r^{\prime \prime
}\right) _{P^{\prime \prime }}\right] \Vert \left[ \left( s^{\prime }\right)
_{m-P^{\prime }},\left( s^{\prime \prime }\right) _{m-P^{\prime \prime
}}\right] }>  \nonumber
\end{eqnarray}
By using of this relation, make the similar states as in Eq.(12), and make
generalization to mixes-states, we can prove that generally, a mixed-state
density matrix $\rho _{i_1\cdots i_N}$ can be reduced through to the
bipartite qubit density matrix $\rho _{\left( \left( r\right) _P-\left(
s\right) _{N-P}\right) }.$ $\square $

The following theorem is the main result in this paper, it is an application
of PPT condition for multipartite qubit systems.

{\bf Theorem 2 (Criterion).}{\it \ For a given partition }$\left( r\right)
_P\Vert \left( s\right) _{N-P}${\it , a necessary condition of a N-partite(N}%
$\geqslant 3)${\it \ qubit density matrix }$\rho _{i_1i_2\cdots i_N}${\it \
to be }$\left( r\right) _P-\left( s\right) _{N-P}${\it -}$separable${\it \
is that the reduced bipartite qubit density matrix is separable, i.e. }$\rho
_{\left( \left( r\right) _P-\left( s\right) _{N-P}\right) \text{ }}$ {\it \ }%
$satisfies${\it \ the PPT condition.}

{\bf Proof. }We only discuss in detail the case of quadripartite qubit, it
can be straightforwardly generalized to the case of arbitrary N-partite
qubit. In the first place, we prove that this theorem holds for a
quadripartite qubit pure-state. Suppose that the pure-state $\rho _{ABCD}$
is AC-BD-separable. This means that if we choose the natural basis $\left\{
\mid i_A>\otimes \mid j_C>\otimes \mid r_B>\otimes \mid s_D>\right\} ,$ then 
$\rho _{AC\Vert BD}=\rho _{AC}\otimes \rho _{BD}$, where $\rho _{AC}=\mid
\Psi _{AC}><\Psi _{AC}\mid ,\;\mid \Psi
_{AC}>=\sum\limits_{i,j=0,1}c_{ij}\mid i_A>\otimes \mid
j_C>,\sum\limits_{i,j=0,1}\left| c_{ij}\right| ^2=1,$ and $\rho _{BD}=\mid
\Psi _{BD}><\Psi _{BD}\mid ,\;\mid \Psi
_{BD}>=\sum\limits_{r,s=0,1}d_{rs}\mid r_B>\otimes \mid s_D>$, $%
\sum\limits_{r,s=0,1}\left| d_{rs}\right| ^2=1.$ From the above ways, it
easily checked that the bipartite qubit density matrix $\rho _{\left(
AC-BD\right) }$, in fact, can be rewritten as 
\begin{eqnarray}
\rho _{\left( AC-BD\right) } &=&\sigma _\vartriangle +\sigma _{\times
}+\sigma _{\diamond }+\sigma _{\wedge }=\sigma _{\left( AC\right) }\otimes
\sigma _{\left( BD\right) }+\sigma _{\left( AC\right) }\otimes \sigma
_{\left( B\stackrel{\vee }{D}\right) }  \nonumber \\
&&+\sigma _{\left( A\stackrel{\vee }{C}\right) }\otimes \sigma _{\left(
BD\right) }+\sigma _{\left( A\stackrel{\vee }{C}\right) }\otimes \sigma
_{\left( B\stackrel{\vee }{D}\right) }
\end{eqnarray}
where $\sigma _{\left( AC\right) }=\mid \Phi _{\left( AC\right) }><\Phi
_{\left( AC\right) }\mid $ , we already write $\mid \Phi _{\left( AC\right)
}>=\sum\limits_{i=0,1}e_i\mid i_x>,e_i\equiv c_{ii}$ and $\mid i_A>\otimes
\mid i_C>\longrightarrow \mid i_x>$. Similarly, $\sigma _{\left( A\stackrel{%
\vee }{C}\right) }=\mid \Phi _{\left( A\stackrel{\vee }{C}\right) }><\Phi
_{\left( A\stackrel{\vee }{C}\right) }\mid ,\ \mid \Phi _{\left( A\stackrel{%
\vee }{C}\right) }>=\sum\limits_{j=0,1}f_j\mid j_x>$, $f_j\equiv c_{j\left(
1-j\right) }$ and $\mid j_B>\otimes \mid \left( 1-j\right)
_D>\longrightarrow \mid j_x>$, and similarly for $\sigma _{\left( BD\right)
},\sigma _{\left( B\stackrel{\vee }{D}\right) },$ etc.. Now, $\rho _{\left(
AC-BD\right) }$ can be written as 
\begin{eqnarray}
\rho _{\left( AC-BD\right) } &=&\eta _{\left( AC\right) }^2\eta _{\left(
BD\right) }^2\rho _{\left( AC\right) }\otimes \rho _{\left( BD\right) }+\eta
_{\left( AC\right) }^2\eta _{\left( B\stackrel{\vee }{D}\right) }^2\rho
_{\left( AC\right) }\otimes \rho _{\left( B\stackrel{\vee }{D}\right) } \\
&&+\eta _{\left( A\stackrel{\vee }{C}\right) }^2\eta _{\left( BD\right)
}^2\rho _{\left( A\stackrel{\vee }{C}\right) }\otimes \rho _{\left(
BD\right) }+\eta _{\left( A\stackrel{\vee }{C}\right) }^2\eta _{\left( B%
\stackrel{\vee }{D}\right) }^2\rho _{\left( A\stackrel{\vee }{C}\right)
}\otimes \rho _{\left( B\stackrel{\vee }{D}\right) }  \nonumber
\end{eqnarray}
where $\rho _{\left( AC\right) }=\left( \eta _{\left( AC\right) }\right)
^{-1}\mid \Phi _{\left( AC\right) }><\Phi _{\left( AC\right) }\mid ,\eta
_{\left( AC\right) }=\sqrt{\sum\limits_{i=0,1}\left| c_{ii}\right| ^2}.$
Now, $\rho _{\left( AC\right) }$ is a density matrix of a single particle.
Similarly, for $\rho _{\left( A\stackrel{\vee }{C}\right) },\rho _{\left(
BD\right) },\rho _{\left( B\stackrel{\vee }{D}\right) }.$ Since 
\begin{eqnarray}
&&\eta _{\left( AC\right) }^2\eta _{\left( BD\right) }^2+\eta _{\left(
AC\right) }^2\eta _{\left( B\stackrel{\vee }{D}\right) }^2+\eta _{\left( A%
\stackrel{\vee }{C}\right) }^2\eta _{\left( BD\right) }^2+\eta _{\left( A%
\stackrel{\vee }{C}\right) }^2\eta _{\left( B\stackrel{\vee }{D}\right) }^2 
\nonumber \\
&=&\left( \eta _{\left( AC\right) }^2+\eta _{\left( A\stackrel{\vee }{C}%
\right) }^2\right) \left( \eta _{\left( BD\right) }^2+\eta _{\left( B%
\stackrel{\vee }{D}\right) }^2\right) =1
\end{eqnarray}
therefore $\rho _{\left( AC-BD\right) }$ is a separable bipartite qubit
mixed-state. The PPT condition for separability of 2$\times 2$ systems is
sufficient and necessary[2], thus $\rho _{\left( AC-BD\right) }$ satisfies\
the PPT condition. Similarly, for other partial separability.

Secondly, we prove that this theorem holds yet for partially separable
mixed-states. Suppose that $\rho _{ABCD}$ is a AC-BD-separable mixed-state,
then under the same natural basis there is a decomposition as $\rho
_{AB\Vert CD}=\sum_\alpha p_\alpha \rho _{\alpha \left( AC\right) }\otimes
\rho _{\alpha \left( BD\right) }$, where $\rho _{\alpha \left( AC\right) }$
and $\rho _{\alpha \left( BD\right) }$ both are bipartite qubit pure-states
as in the above for all $\alpha ,$ $0<p_\alpha \leq 1,$ $\sum\limits_\alpha
p_\alpha =1.$ From the above reduction operation, obviously we have 
\begin{equation}
\rho _{\left( AC-BD\right) }=\sum_\alpha p_\alpha \left[ \rho _{\alpha
\left( AC\right) }\otimes \rho _{\alpha \left( BD\right) }\right] _{\left(
AC-BD\right) }
\end{equation}
According to the above mention, every $\left[ \rho _{\alpha \left( AC\right)
}\otimes \rho _{\alpha \left( BD\right) }\right] _{\left( AC-BD\right) }$ is
a separable bipartite qubit mixed-state, this leads to that the convex sum $%
\rho _{\left( AC-BD\right) }$ in Eq.(20) still is a separable bipartite
qubit mixed-state, and it must satisfy\ the PPT condition.

Similarly, we cane prove higher dimensional cases. $\square $

{\bf Corollary. }{\it If the reduced bipartite\ qubit density matrix }$%
\left( \rho _{i_1i_2\cdots i_N}\right) _{\left( \left( r\right) _P-\left(
s\right) _{N-P}\right) }${\it \ violates the PPT condition for a partition }$%
\left( r\right) _P\Vert \left( s\right) _{N-P}${\it , then }$\rho
_{i_1i_2\cdots i_N}${\it \ is }$\left( r\right) _P-\left( s\right) _{N-P}$%
{\it -inseparable and} {\it entangled.}

It, in fact, is the inverse-negative proposition of Theorem 2.

{\bf Examples. }Consider two{\bf \ }tripartite qubit density matrixes 
\begin{eqnarray}
\rho _{ABC}^{\prime } &=&\left[ 
\begin{array}{cccccccc}
0 &  &  &  &  &  &  &  \\ 
& \frac{1-x}4 &  &  &  &  &  &  \\ 
&  & \frac{1-x}4 &  &  &  &  &  \\ 
&  &  & \frac x2 & -\frac x2 &  &  &  \\ 
&  &  & -\frac x2 & \frac x2 &  &  &  \\ 
&  &  &  &  & \frac{1-x}4 &  &  \\ 
&  &  &  &  &  & \frac{1-x}4 &  \\ 
&  &  &  &  &  &  & 0
\end{array}
\right]  \nonumber \\
\rho _{ABC}^{\prime \prime } &=&\left[ 
\begin{array}{cccccccc}
0 &  &  &  &  &  &  &  \\ 
& \frac{1-x}4 &  &  &  &  &  &  \\ 
&  & \frac x2 & 0 & 0 & -\frac x2 &  &  \\ 
&  & 0 & \frac{1-x}4 & 0 & 0 &  &  \\ 
&  & 0 & 0 & \frac{1-x}4 & 0 &  &  \\ 
&  & -\frac x2 & 0 & 0 & \frac x2 &  &  \\ 
&  &  &  &  &  & \frac{1-x}4 &  \\ 
&  &  &  &  &  &  & 0
\end{array}
\right]
\end{eqnarray}
then we have 
\begin{equation}
\left( \rho _{ABC}^{\prime }\right) _{\left( A-BC\right) }=\left( \;\rho
_{ABC}^{\prime \prime }\right) _{\left( B-AC\right) }=\rho _W
\end{equation}
where $\rho _W$ is the Werner state[1,15$]$ which consists of a singlet
fraction $x$ and a random fraction $(1-x)$, 
\begin{eqnarray}
\left[ \rho _W\right] _{ij,rs} &=&xS_{ij,rs}+\frac 14\left( 1-x\right)
\delta _{ir}\delta _{js}  \nonumber \\
S_{01,01} &=&S_{10,10}=-S_{01,10}=-S_{10,01}=\frac 12 \\
&&\text{and all the other components of }S\text{ vanish}.  \nonumber
\end{eqnarray}
It is known[1$]$ that when $\frac 13<x\leq 1$ $\rho _W$ violates the PPT
condition, it leads to that $\rho _{ABC}^{^{\prime }}$ is A-BC-inseparable
and $\rho _{ABC}^{\prime \prime }$ is B-AC-inseparable.

By using of the above theorems and corollary, in some special cases we can
make a $N$-partite qubit from $2^{N-2}$ bipartite qubit density matrixes,
which is partially inseparable for a given partition. As in the above, for
the case of tripartite qubit we take two bipartite qubit density matrixes $%
\sigma _{\left( 1\right) },$ $\sigma _{\left( 2\right) }$ and real numbers $%
p_1,$ $p_2,\;0<p_1,$ $p_2\leq 1$ such that $\sigma =p_1\sigma _{\left(
1\right) }+$ $p_2$ $\sigma _{\left( 2\right) }$ is a bipartite qubit
entangled state ( then it violates the PPT condition). If we want to
construct a tripartite qubit entangled state $\rho _{ABC}$ which is
B-AC-inseparable, then we can take the entries of $\rho _{ABC}$ by 
\begin{eqnarray}
\left[ \rho _{ABC}\right] _{ijk,rst} &=&p_1\left[ \sigma _{\left( 1\right)
}\right] _{ji,sr},\text{ for }k=i\text{ and }t=r\;  \nonumber \\
\left[ \rho _{ABC}\right] _{ijk,rst} &=&p_2\left[ \sigma _{\left( 2\right)
}\right] _{ji,sr},\text{ for }k=1-i\text{ and }t=1-r \\
\left[ \rho _{ABC}\right] _{ijk,rst} &=&0,\text{ for the rest }%
(i,j,k,r,s,t=0,1)  \nonumber
\end{eqnarray}
It can be verified that $\rho _{ABC}$ is a tripartite qubit density matrix,
and is B-AC-inseparable. In fact, $\left( \rho _{ABC}\right) _{\left(
B-AC\right) }=\tau $ which violates the PPT condition. Similarly, for A-BC
and C-AB. The above way can be generalized to obtain a $\left( r\right)
_P-\left( s\right) _{N-P}$-inseparable density matrix from a bipartite qubit
entangled state in form as $\tau =\sum\limits_{i=1}^{2^{N-2}}p_i\sigma
_{\left( i\right) },$ where all $\sigma _{\left( i\right) }$ are some
bipartite qubit density matrixes.

\_\_\_\_\_\_\_\_\_\_\_\_\_\_\_\_\_\_\_\_\_\_\_\_\_\_\_\_\_\_\_\_\_\_\_\_\_\_%
\_\_\_\_\_\_\_\_\_\_\_\_\_\_\_\_\_\_\_\_\_\_\_\_\_\_\_\_\_\_\_\_\_\_\_\_\_\_%
\_\_\_\_\_-

\bigskip

\bigskip

{\Large References}

[1] A. Peres, Phys. Rev. Lett., {\bf 77}(1996)1413.

[2] M. Horodecki, P. Horodecki, and R. Horodecki, Phys. Lett. A,

\ \ \ \ {\bf 223}(1996)1.

[3] S. Wu. X. Chen, and Y. Zhang, Phys. Lett. A, {\bf 275}(2000)244.

[4] M. Horodecki, P. Horodecki, and R. Horodecki,

\ \ \ \ quant-ph/0006071.

[5] B. M. Terhal, J. Theo. Compu. Sci., {\bf 287}(1)(2002)313.

[6] M. Horodecki, P. Horodecki, and R. Horodecki,

\ \ \ \ quant-ph/0206008.

[7] K. Chen, and L. A. Wu, Phys. Lett. A, {\bf 306}(2002)14.

[8] W. D\"{u}r, G. Vidal, and J. I. Cirac, Phys. Rev. A, {\bf 62}%
(2000)062314.

[9] M. Seevinck and G. Svetlichny, Phys. Rev. Lett., {\bf 89}(2002)060401.

[10] A. O. Pittenger and M. H. Rubin, Phys. Rev. A, {\bf 62}(2000)032313.

[11] A. M. Wang, quant-ph/0305016.

[12] T. Yamakami, quant-ph/0308072.

[13] C. H. Bennett , D. P. DiVincenzo , T. Mor, P. W. Shor,

\ \ \ \ \ J. A. Smolin, and B. M. Terhal, Phys. Rev. Lett., {\bf 82}%
(1999)5385.

[14] D. P. DiVincenzo , T. Mor, P. W. Shor, J. A. Smolin,

\ \ \ \ \ and B. M. Terhal, Comm. Math. Phys., {\bf 238}(2003)379.

[15] J. Blank and P. Exner, Acta Univ. Carolinae, Math. Phys. 18(1977)3.

\end{document}